\documentclass[runningheads]{llncs}
\usepackage{color}
\usepackage{tabu} 
\usepackage[pdftex]{graphicx} 
\usepackage{booktabs,floatrow}
\usepackage{pifont}

\newcommand{\myetal}{\textit{et al.}}
\newcommand{\myie}{\textit{i.e.,}}
\newcommand{\myeg}{\textit{e.g.,}}
\newcommand{\myetc}{\textit{etc.}}

\newcommand{\mytool}{Astra}

\definecolor{eclipseBlue}{RGB}{42,0.0,255} 
\definecolor{eclipseGreen}{RGB}{63,127,95} 
\definecolor{eclipsePurple}{RGB}{127,0,85} 
\definecolor{eclipseCoral}{RGB}{255,127,80} 
\definecolor{eclipseBrown}{RGB}{40,26,13}

\newcommand{\trans}[1]{[#1]}
\newcommand{\typ}[1]{\mbox{\textsc{type}}(#1)}

\newcommand{\fcntype}{\rightarrow}
\usepackage[formats]{listings} % Define Language 
\lstdefinelanguage{alloy} { 	% list of keywords
    morekeywords={
         abstract,
         all,
         and,
         as,
         assert,
         but,
         check,
         disj,
         else,
         exactly,
         extends,
         fact,
         for,
         fun,
         iden,
         iff,
         implies,
         in,
         Int,
         let,
         lone,
         module,
         no,
         none,
         not,
         one,
         open,
         or,
         pred,
         run,
         set,
         sig,
         some,
         sum,
         univ
    }, 	
    sensitive=true, % keywords are case-sensitive
    morecomment=[l]{//}, % l is for line comment 	
    morecomment=[s]{/*}{*/}, % s is for start and end delimiter 	
    morestring=[b]" % defines that strings are enclosed in double quotes 
    } % Set Language 
    \lstdefinestyle{alloy}{ 	
    language={alloy}, 	
    basicstyle=\footnotesize\ttfamily, % Global Code Style 	
    captionpos=b, % Position of the Caption (t for top, b for bottom) 
    extendedchars=true, % Allows 256 instead of 128 ASCII characters 	
    tabsize=2, % number of spaces indented when discovering a tab 	
    columns=fixed, % make all characters equal width 	
    keepspaces=true, % does not ignore spaces to fit width, convert tabs to spaces
    showstringspaces=false, % lets spaces in strings appear as real spaces
    breaklines=true, % wrap lines if they don't fit 	
    framesep=4pt, % quarter circle size of the round corners 	
    numbers=left, % show line numbers at the left  
    numberstyle=\footnotesize\ttfamily, % style of the line numbers 
    commentstyle=\color{eclipseGreen}, % style of comments 
    keywordstyle=\bfseries\color{eclipsePurple}, % style of keywords 
    stringstyle=\color{eclipseBlue}, % style of strings 	
    xleftmargin=0.8cm, 	escapeinside={\%*}{*)}          % if you want to add LaTeX within your code 
} 
\newcommand{\alloycode}[1]{\lstinline[style=alloy]{#1}}

\begin{document}
\title{\mytool\ Version 1.0: Evaluating Translations\\ from Alloy to SMT-LIB}
\author{Ali Abbassi\inst{1} \and
Nancy A. Day\inst{1} \and
Derek Rayside\inst{2}
\authorrunning{A. Abbassi et al.}
\institute{David R. Cheriton School of Computer Science, University of Waterloo, Waterloo, ON, Canada 
\email{\{aabbassi,nday\}@uwaterloo.ca} \and
Department of Electrical and Computer Engineering, University of Waterloo, Waterloo, ON, Canada 
\email{drayside}@uwaterloo.ca}}

\maketitle              

\begin{abstract}
We present a variety of translation options
for converting Alloy to SMT-LIB via Alloy's 
Kodkod interface.
Our translations, which are implemented in a library that we call Astra, 
are based on converting 
the set and relational operations 
of Alloy into their equivalent in typed first-order logic (TFOL).  
We investigate and compare the performance of an SMT solver for many translation options.  
We compare using only one universal type
to recovering Alloy type information from the Kodkod representation 
and using multiple types in TFOL.
We compare a direct translation of the relations to predicates
in TFOL to one where we recover functions from their
relational form in Kodkod and represent these as functions in TFOL.  
  We compare representations in TFOL with
unbounded scopes to ones with bounded scopes, either pre or post
quantifier expansion.  Our results across all these dimensions
provide directions for portfolio solvers, modelling 
improvements, and optimizing SMT solvers.

\end{abstract}

\section{Introduction}

The Alloy Analyzer~\cite{Ja2012alloy} is a well-used formal analysis tool because of 
its high-level modelling language based on first-order logic and sets, and because of 
its convenient tool support.  Its tool support is easy-to-use in part because its analysis is 
fully automated, which is possible because it checks only for sets of finite scope.  
Finite model finding provides users with quick results for small size problems 
when debugging their models.  However, as verification problems become more detailed 
or the scopes become larger, the analysis engine of Alloy called Kodkod~\cite{ToJa2007tacas} 
reaches performance limitations.
The Kodkod engine treats all predicates and functions as relations and 
implements Alloy's typing system via predicates.  Kodkod expands quantified 
formulas over finite domains and then passes the problem to a SAT solver.

SMT (satisfiability modulo theories) solvers~\cite{barret-smt09} continue to 
improve in performance.  These are solvers for first-order logic (FOL) formulas 
enriched with decision procedures for specific datatypes such as equality and linear arithmetic. 
Even for formulas outside of a decidable subset of FOL, an SMT solver can often reach a 
conclusion of either a satisfying solution or that no solution exists.

In this paper, we investigate the performance of number of translation options for
converting Alloy to typed FOL (TFOL).  We call our library
Astra (\underline{A}lloy to \underline{S}MT-LIB \underline{tra}nslation).
To enable future integration with the Alloy Analyzer, we start from the Kodkod interface.
We compare using only one universal type
to recovering Alloy type information from the Kodkod representation 
and using multiple types in TFOL.
We compare a direct translation of the relations to predicates
in TFOL to one where we recover functions from their
relational form in Kodkod and represent these as functions in TFOL.  
  We compare representations in TFOL with
unbounded scopes to ones with bounded scopes, either pre or post
quantifier expansion. 
TFOL is represented as formulas in the SMT-LIB format.
We run our performance tests using the SMT solver Z3~\cite{MoBj2008tacas} to compare the
results of all of our translation options to each other and Kodkod.
We categorized our tests by characteristics of the model such as 
depth of quantifiers, number of types, the use of join, 
number of functions, \myetc\ and discuss whether there are any
meaningful correlations between model characteristics and 
our translation options.

There are several previous efforts at translating Alloy to 
SMT solvers or other automated solvers (\myeg\ ~\cite{elghazi-fm11,elghazi-corr-15,ulbrich-assistant})
for the purposes of analyzing Alloy models of unbounded scope particularly
for built-in Alloy types.  Our translation options cover some of this
work, however, none of the previous efforts compare different translation options
or try to correlate them with model characteristics.  
Also, we start from the Kodkod
interface for ease of future integration with the Alloy Analyzer.  
There have also been efforts to create theories for 
finite model finding in SMT solvers (\myeg~\cite{bansal-corr17,reynolds-cav13}),
but these have not yet been linked with Alloy.

Our results are useful to anyone who uses Alloy for verification.  
We have not yet implemented transitive closure, cardinality of sets,
or support for built-in Alloy types.
Our contribution comparing all of these translation options
provides directions for portfolio solvers, modelling 
improvements, and optimizing SMT solvers.

\section{Background}

\textbf{Alloy.}
Figure~\ref{fig:alloy-example} shows an
example of an Alloy model that we use for illustration 
throughout the paper. Signatures are declarations that 
create a set.  Within the signature, relations from that set 
to other sets are declared, such as on line~\ref{relation-line},
which declares a relation from elements in the set \alloycode{A} to elements
in the set \alloycode{ID}.
These relations can have cardinality constraints within their
declarations.  For example, on line~\ref{relation-line}, the range
has the keyword \alloycode{one}, which means that \alloycode{id}
is a total function, \myie\ each domain element is associated with exactly
one range element.  Alternative cardinality constraints are \alloycode{lone},
which means it is a partial function and \alloycode{set}, which means it
is a relation. 
Constraints in Alloy are formulas
in FOL and set theory, such as line~\ref{fact-line}, which forces
\alloycode{id} to be an injective function.

Alloy provides
type inheritance where one signature can be a subset of another.
In our example, \alloycode{B} and \alloycode{C} are subtypes of \alloycode{A}
(using the keyword \alloycode{extends}).
Since set \alloycode{A} is declared as abstract, there are no elements in \alloycode{A}
that are not in either \alloycode{B}
or \alloycode{C}. The sets \alloycode{B} and \alloycode{C} are disjoint.
For a signature that is not declared as abstract, there can be elements in the
parent set that are not in the child sets.

Alloy models make frequent use of the join operator between relations
since everything in Alloy is a set. For example on line~\ref{fact-line},
\alloycode{a.id} is the join of the singleton set consisting of \alloycode{a}
with the \alloycode{id} relation, resulting in any range elements of \alloycode{id}
that have \alloycode{a} as their first element.

Using the Alloy Analyzer, 
a user can set finite scopes for the sets
of the model and check their model using Alloy's finite model finding
analysis implemented in Kodkod.
A \alloycode{run} command (as on lines~\ref{run-line-start}-\ref{run-line-end}) asks the Alloy Analyzer
to produce an instance of these sets and relations that satisfy the constraints.
Using a \alloycode{check} command, the Analyzer will determine if an assertion
holds of the model.

\begin{figure*}
\begin{lstlisting}[style=alloy]
sig ID {}
abstract sig A {
    id: one ID %* \label{relation-line}  *)
}
sig B extends A {
    toC: one C
}
sig C extends A {
    toB: set B
}
fact id_is_injective {
    all a, a': A | a.id = a'.id => a = a' %* \label{fact-line}  *)
}

run {} for exactly 6 A, exactly 3 B, %* \label{run-line-start}  *)
           exactly 3 C, exactly 6 ID %* \label{run-line-end}  *)
\end{lstlisting}
\caption{Example of an Alloy model.}
\label{fig:alloy-example}
\end{figure*}

\textbf{Kodkod.} Kodkod~\cite{ToJa2007tacas} is the component of the Alloy Analyzer
that implements its finite model finding.  An Alloy model is
converted to a representation as a datatype in Kodkod.
Kodkod supports only relations -- there are no types and no
functions.  Types are represented as unary relations.  Functions
and relations with multiplicity constraints are all treated as 
relations with extra constraints to represent their multiplicity restrictions.
Kodkod supports propositional logic operations, universal and
existential quantification, and a set of common operators on sets including
join.
Kodkod is a Java library
and the Alloy Analyzer can produce a Java program that creates the
Kodkod representation of an Alloy model.  This code has the following
parts that are of interest to us: 1) an atomlist, a list of constants 
that represent values in the finite sets of the model;
2) a data structure called bounds, which populates each relation of
the model with its possible tuples from the constants;
and 3) a data structure representing the constraints of the Alloy
model together with the problem to be solved (from the \alloycode{run} or
\alloycode{check} commands) as one conjunction.
Kodkod performs quantifier expansion, simplifications and symmetry reductions to produce
a problem for a SAT solver.

\textbf{Fortress.} 
One of our translation options is to use the existing library
Fortress to represent models of finite scope.
Fortress is a Java library created by Vakili and Day~\cite{VaDa16fm}
that reduces the finite model finding problem for FOL 
to the logic of equality with uninterpreted
functions (EUF).  The idea is that an EUF solver
can exploit
the structure of functions and types used in the model
to achieve better performance than Kodkod and other related
finite model finding methods.
Fortress introduces
constants for the elements of the finite scope.  Then, it creates
range formulas, which are constraints on the outputs of each function
to be one of the elements of the scope, which force
the EUF problem to have solutions only within a
finite scope.  Fortress performs quantifier expansion, simplifications,
and symmetry reductions before producing a problem for the SMT
solver Z3~\cite{MoBj2008tacas}.  Vakili and Day~\cite{VaDa16fm} show
Fortress has quite good results compared to Kodkod on a number
of problems.  We use Fortress' abstract datatype for TFOL in
our implementation.

\textbf{SMT-LIB2 and SMT Solvers.}
SMT solvers~\cite{barret-smt09} search for satisfying instances of FOL problems by using
cooperating decision procedures for particular subsets 
of FOL that have standard
interpretations (such as equality or linear arithmetic).  The
standard interface language for SMT solvers is SMT-LIB2~\cite{clark2015},
which we use in this work.
We use the SMT solver Z3~\cite{MoBj2008tacas} although other
solvers such as CVC4~\cite{barret-cav11} can be used.

\section{Translating Kodkod to Typed First-Order Logic (TFOL)}

\label{sec:trans}

\begin{figure*}
\includegraphics[width=9cm]{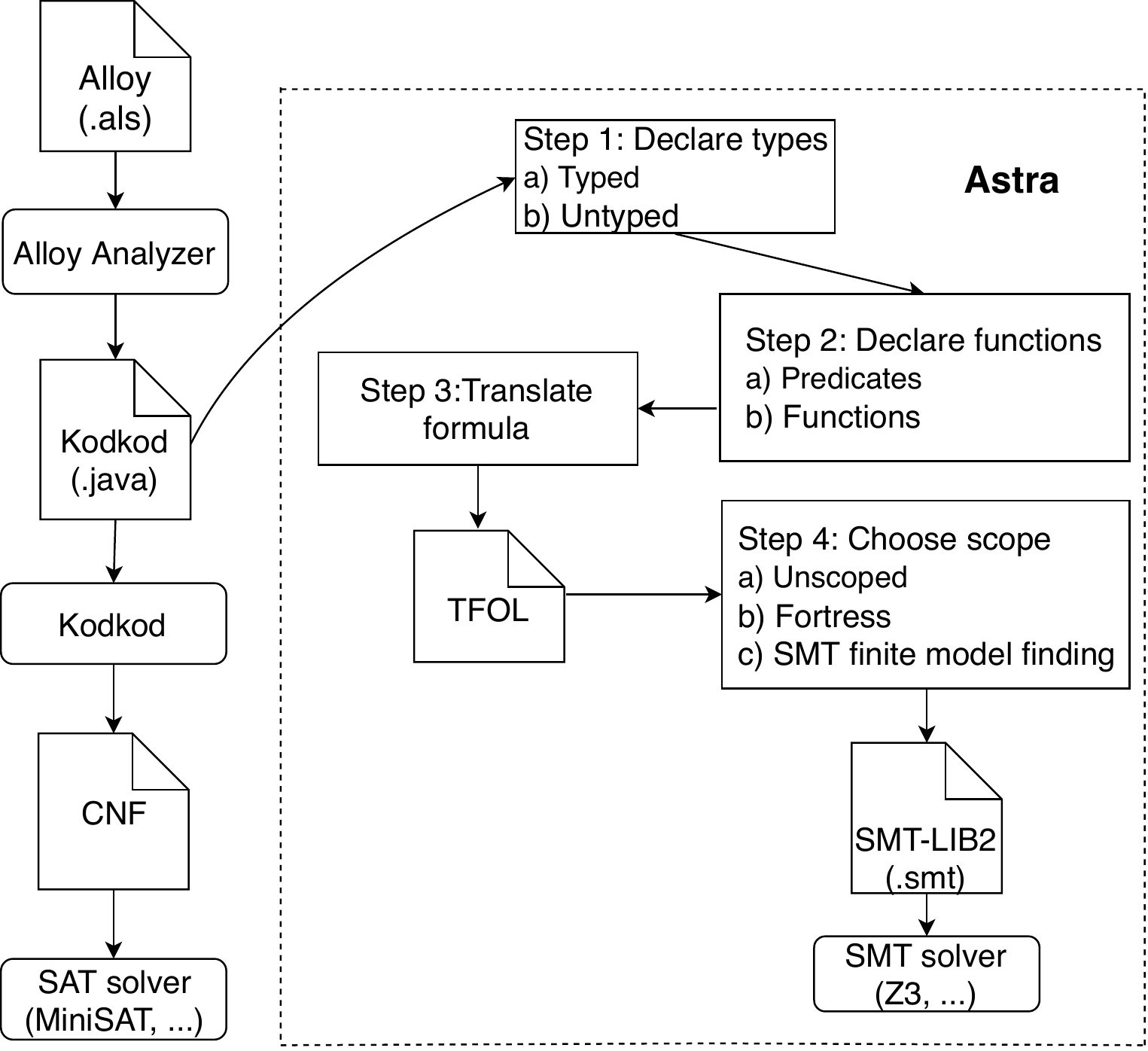}
\caption{Steps in translation.}
\label{fig:trans}
\end{figure*}

Figure~\ref{fig:trans} presents an outline of
the steps we use to translate Kodkod to TFOL.
TFOL supports only total functions.
Our goal is to create a very direct translation to
TFOL from Alloy.
Our translation is similar in parts to Ulbrich \myetal~\cite{ulbrich-assistant}'s
translation of Alloy to KeY~\cite{key} and El Ghazi~\myetal~\cite{elghazi-fm11}'s translation
of Alloy to SMT although both of those efforts implement only one combination
of our options.
Next, we describe the steps in our translation and the 
options we investigate for each step.  To do the translation,
we work from Kodkod's atomlist, bounds, and formula data structures.

\textbf{Step 1: Declaring Types.}
Alloy uses sets as types.  In the Kodkod representation,
there is one universe of elements and unary relations constrain
the elements to be of a certain type when needed.  In TFOL, 
we have the ability to use its type space to separate
the elements into types and thereby reduce the size of the set of 
elements for each quantifier.  \textit{We hypothesize that using types
will improve the performance of the SMT solver.}
Our first translation step is to declare types in TFOL. We
expore two options: untyped and typed.

For the untyped option,
we introduce one ``universal'' type (called ``Univ'' in our
examples in the paper) to use as the types
of all functions in TFOL. 

In the typed option, there are various levels in the
Alloy type hierarchy we could choose as FOL types.
To investigate an opposite translation point from
untyped, we choose the types at the leaves of the hierarchy as
the TFOL types.   For the example of Figure~\ref{fig:alloy-example},
the types declared are \alloycode{B} and \alloycode{C}.

Kodkod does not store the type hierarchy of the Alloy model
directly.  We have to reverse engineer it from the atomlist
and bounds data structures.  Kodkod names its constants
prefixing them with type names of the leafs of the
type hierarchy.  This information tells us the names of the
leaf types so we can declare them in TFOL.  Some of these
type names are for a set of elements that are
not part of a declared leaf type in Alloy because the subtypes
do not necessarily include all elements of the parent type.

\textbf{Step 2: Declaring Functions.}
The next step of translation is to declare the total functions
in TFOL. A predicate
in TFOL is a function that returns a Boolean value.

In this step, we traverse Kodkod's 
bounds datatype, ignoring unary relations that were used as
types if the typed option was chosen for Step 1.  
For the remaining relations,
we create function declarations in TFOL. 
\textit{We hypothesize that solving a problem with theories is harder
than solving a SAT problem in an SMT solver. Therefore, using predicates instead of
functions can improve the performance, at least for UNSAT problems,
since if the SMT solver cannot find a satisfying Boolean assignment,
then it does not have to use its theories.}
Thus, we implement two options for declaring TFOL functions.
The first, which we call ``predicates'', is to make all Kodkod relations 
predicates in TFOL.  The
second, which we call ``functions'', is to use as 
many non-predicate functions as possible in TFOL.  

For the predicates option, we declare the predicates in TFOL
from the information in the bounds datatype directly.  For the untyped
option of Step 1, we declare all relations to take arguments of
the universal type and return a Boolean.  
The  
TFOL function declarations for the type hierarchy
of Figure~\ref{fig:alloy-example} for the untyped option are:
\begin{center}
\begin{math}
\begin{array}{l}
id: Univ \times Univ \fcntype Bool \\
toC: Univ \times Univ \fcntype Bool \\
toB: Univ \times Univ \fcntype Bool \\
\end{array}
\end{math}
\end{center}
We then add constraints to describe the multiplicity restrictions.

For the functions option, we have to 
reverse engineer from the Kodkod datatypes which of these relations 
are total functions in Alloy.  We traverse the
Kodkod formula looking for multiplicity constraints.
The information that a relation is total function is found as a constraint
of a certain pattern in the formula data structure of Kodkod indicating the
range of the function is \alloycode{one X}, which means we can declare
it as having a return type X in TFOL. For example, the TFOL declaration of the total
function
\alloycode{toC} in Figure~\ref{fig:alloy-example} for the functions option
would be:
$$toC: Univ \fcntype Univ$$
and fewer multiplicity restrictions are needed.
For relations that have non-total
function multiplicities (including \alloycode{lone} multiplicities, which
means it is a partial function), we declare them as predicates of the appropriate
type and add their multiplicity constraints as formulas in our TFOL model.
When we translate the formula in Step 3, we ignore these multiplicity
constraints in Kodkod because they have already been translated.

For the typed option of Step 1, the TFOL function declarations are
more complicated.
We read the leaf type names from the names of constants in
the tuples Kodkod assigned to the relation.
For relations that have domain
and range types that are non-leaf types, we create
multiple copies of the relation.  For example, in
Figure~\ref{fig:alloy-example}, 
\texttt{id} is a function which has a domain that is a non-leaf type, 
\texttt{A},
thus the TFOL declarations for the predicates option are:
\begin{center}
\begin{math}
\begin{array}{l}
id1: B \times ID \fcntype Bool \\
id2: C \times ID \fcntype Bool \\
toC: B \times C \fcntype Bool \\
toB: C \times B \fcntype Bool \\
\end{array}
\end{math}
\end{center}
For the functions option, the TFOL declarations are:
\begin{center}
\begin{math}
\begin{array}{l}
     id1: B \fcntype ID\\
     id2: C \fcntype ID \\
     toC: B \fcntype C \\
     toB: C \times B \fcntype Bool \\
\end{array}
\end{math}
\end{center}

\textbf{Step 3: Translating Formulas.}
Step 3 is to translate the formulas in the Kodkod datatype to formulas in TFOL.
In this step, we have to translate set operations (union, intersection, \myetc)
into their equivalent in FOL.  
Our translation is defined by the $[\cdot]$ operator,
which takes a formula in Kodkod, and translates it into a TFOL formula.
We define $[\cdot]$ in this section.

To create as direct a translation to TFOL as possible, 
we represent each set operation using the characteristic
predicate for the set and the propositional operation that
is the equivalent of the set operation.  For example, 
the union of two sets is the disjunction of the characteristic
predicate for each of the operands to the union.

This translation can be done in either a top-down or bottom-up
traversal of the Kodkod formula data structure.  To handle the
generality of set expressions in Alloy, we choose
a bottom-up traversal to facilitate its compositionality 
and so that the types of terms can be determined
from their leaves on the way up.  To make a bottom-up traversal
possible, we have to provide a translation for each term, not
just each formula.  We describe our translation below for the typed
option in Step 1.  

The leaves of the Kodkod formula datatype are variables or relations of certain types.
We need this type information in our translation to TFOL.  Kodkod stores the
types of its variables with the variable, so the translation of a 
Kodkod variable or relation is simply a term of the type in the Kodkod, as in
$\trans{v:t} := v:t$.

Next, we describe the translation for the set operations.  
A term in Kodkod must return a term in TFOL
so we translate each non-leaf term into a helper relation
and add a constraint for the meaning of the helper relation.
For example, $\trans{A \cup B}$ where $A$ and $B$ are both of
type $t$ is a new relation $R$ of type $t \rightarrow Bool$
with an additional constraint of 
$\forall x:t \:\:\bullet\:\: R(x) \Leftrightarrow A(x) \lor B(x)$.
Thus, the translation of the set operations results in a TFOL term
plus declarations and additional constraints.    The type of the
Kodkod term must be determined as we walk up the data structure; 
we use the notation $\typ{\cdot}$ to denote this calculation.
In the following, each $R_{new}$ is a new relation name:

\begin{center}
\begin{math}
\begin{array}{rcl}
    \trans{A \cup B} & :=  & R_{new}:\typ{A} \fcntype Bool  \\
                     &     & \mbox{add}\: \forall x:\typ{A} \bullet R_{new}(x) \Leftrightarrow \trans{A}(x) \lor \trans{B}(x) \\
    \trans{A \cap B} & :=  & R_{new}:\typ{A} \fcntype Bool   \\
                     &     & \mbox{add}\: \forall x:\typ{A} \bullet R_{new}(x) \Leftrightarrow \trans{A}(x) \land \trans{B}(x) \\
    \trans{A - B} & :=  & R_{new}:\typ{A} \fcntype Bool   \\
                     &     & \mbox{add}\: \forall x:\typ{A} \bullet R_{new}(x) \Leftrightarrow \trans{A}(x) \land \neg\trans{B}(x) \\
    \trans{^\sim A}  & :=  & R_{new}:\typ{ran(A)} \times \typ{dom(A)} \fcntype Bool   \\
                     &     & \mbox{add}\: \forall x:\typ{ran(A)}, y:\typ{dom(A)} \bullet \\
                     &     & \hspace*{0.5cm} R_{new}(x,y) \Leftrightarrow \trans{A}(y,x)  \\
    \trans{((v_1:t_1), (v_2:t_2))} & := & v_3:t_1 \times t_2
\end{array}
\end{math}
\end{center}

By using this method of helper relations (also used in~\cite{ulbrich-assistant}, although not for a link with SMT solvers),
our translation results in smaller clauses but more of them.  This method reduces the size
of formulas after quantifier expansion for finite model finding.

Using this method, the general case of translating the very common join operation in Alloy is:

\begin{center}
\begin{math}
\begin{array}{rcl}
    \trans{A . B} & :=  & R_{new}:\typ{dom(A)} \times \typ{ran(B)} \fcntype Bool  \\
                     &     & \mbox{add}\: \forall x:\typ{dom(A)}, y:\typ{ran(B)} \bullet \\
                     & & \hspace{0.5cm} R_{new}(x,y) \Leftrightarrow \exists z:\typ{ran(A)} \bullet \trans{A}(x,z) \land \trans{B}(z,y) \\
\end{array}
\end{math}
\end{center}

For the functions option of Step 2, there are special cases for join where
the first argument, $v$, to join is
a variable or the first argument is the result of a total function application
(which results in a scalar), we can translate this expression to function application:

\begin{center}
\begin{math}
\begin{array}{rcl}
    \trans{(v:t) . f} & :=  &  (f(v:t)):\typ{ran(f)} \\
    \trans{(v:t).f_1.f_2} & := & (f_2(f_1(v:t))):\typ{ran(f_2)}
\end{array}
\end{math}
\end{center}

The remaining formulas of Kodkod (Boolean operations and equality, quantification, $\subseteq$, and $\in$)
have straightforward translations:
\begin{center}
\begin{math}
\begin{array}{rcl}
    \trans{true} & := & true \\
    \trans{false} & := & false \\
    \trans{\neg A} & := & \neg \trans{A} \\
    \trans{A \land B} & :=  & \trans{A} \land \trans{B} \\
    \trans{A \lor B} & := & \trans{A} \lor \trans{B} \\ 
    \trans{A \Rightarrow B} & := & \trans{A} \Rightarrow \trans{B} \\
    \trans{A \Leftrightarrow B} & := & \trans{A} \Leftrightarrow \trans{B} \\
    \trans{A \;=\; B} & := & \trans{A} \;=\; \trans{B} \\ 
    \trans{\forall (x:t) \bullet A} & := & \forall x:t \bullet \trans{A}  \\
    \trans{\exists (x:t) \bullet A} & := & \exists x:t \bullet \trans{A}  \\
    \trans{A \subseteq B} & := & \forall x:\typ{A} \bullet \trans{A}(x) \Rightarrow \trans{B}(x)  \\
    \trans{(v:t) \in B} & := & \trans{B}(v:t)
\end{array}
\end{math}
\end{center}
Everywhere that a relation of non-leaf domain type is used in a formula,
it has to be replaced
by an appropriate operation over the leaf copies of the relation.

If the untyped option is chosen in Step 1, this translation changes slightly.
The types all become the universal type and unary predicates are added 
as appropriate to limit the formula to the correct type.  For example, the
translation for the union operator is:
\begin{center}
\begin{math}
\begin{array}{rcl}
    \trans{A \cup B} & :=  & R_{new}: Univ \fcntype Bool  \\
                     &     & \mbox{add}\: \forall x \bullet P_{\typ{A}}(x) \Rightarrow (R_{new}(x) \Leftrightarrow \trans{A}(x) \lor \trans{B}(x)) \\
\end{array}
\end{math}
\end{center}     
where $P_{\typ{A}}(x)$ is the predicate for the type of set $A$.
The change also affects the translation of the multiplicity constraints in Step 2
in a similar manner.

The alternative of a top-down traversal would have been more difficult
to correctly implement. It would result in longer formulas, but no extra
quantified constraints.  In a top-down traversal of a nested set expression,
variables of unknown type would have had to be created on the way down
the traversal so that formulas would always be passed back up the
traversal.  

\textbf{Step 4: Choosing Scopes.}
In step 4, the scopes for the problem are set.  We can determine
the scopes the user set for each of the Alloy sets by counting
the number of constants with a prefix of a set name in
Kodkod's atomlist.
For the untyped option, the scope is the length of Kodkod's atomlist.

\textit{We hypothesize that if the problem is within a decidable fragment
of TFOL, the SMT solver can solve it quickly without any bounds since the
formula is smaller without introduced constants and their expansion.}
We also want to investigate whether quantifier expansion with finite scopes
is more efficiently done by the SMT solver or prior to solving.
We investigate three methods for handling the scope. 

Our first option is called unscoped
because we leave the types unbounded.
This SMT problem may not be decidable, but it is possible that we will get a result 
from the SMT solver.  

Our second option is called Fortress because we use the existing 
Fortress library. Along with the formula, we need to pass to Fortress
scopes for each type.
Fortress first creates distinct constants for each element of the scope.
Next, it creates range formulas for each function to constrain
it to return one of the constants of the type.
Finally, Fortress expands each quantifier, does symmetry reductions and
simplifications before creating an SMT-LIB2 problem.

We call our third option SMT finite model finding (SMT FMF), where we let the SMT
solver do any needed quantifier expansion over the constants and simplifications.
We declare the constants and create range formulas and pass the problem 
and the range formulas
directly to the SMT solver.

\textbf{Step 5: SMT Solver.}
In the final step, Step 5, the SMT solver is called on the problem.

We have not yet proven the correctness of our translation, however, since the
translation is from Kodkod (rather than Alloy) the number of cases to consider
for correctness is reduced and can be seen more directly.

\textbf{Implementation.} \mytool\ is implemented in Java as a solver
that takes Kodkod's data structures as arguments to facilitate easy future integration
with the Alloy Analyzer.  We use the Fortress library implemented
by Vakili and Day~\cite{VaDa16fm} as our abstract datatype for TFOL declarations
and formulas.

\section{Evaluation}
%======================================================================
Our goal in this work is to investigate whether alternative methods
to Kodkod have better performance.  We expect that no one
method will always have better performance, but perhaps we can
learn which solving method works better for models of certain
characteristics.

\subsection{Tests}

\begin{figure}
\includegraphics[width=\linewidth]{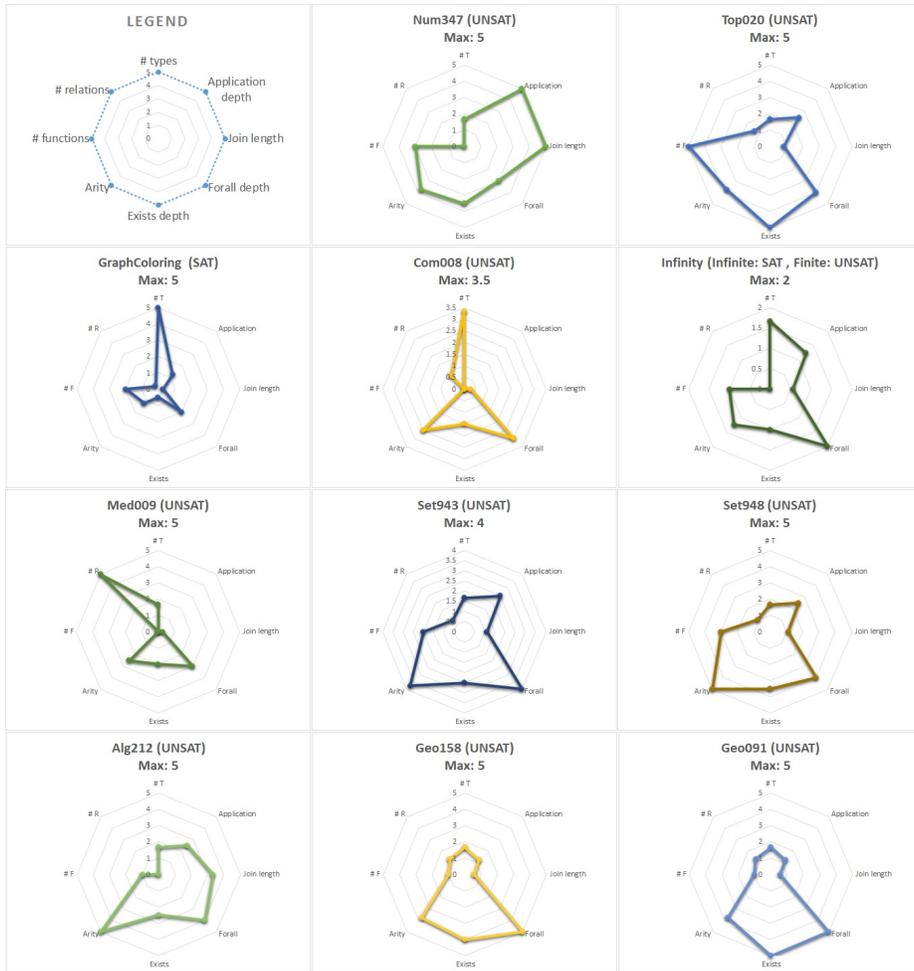}
\caption{Characteristics of tests.}
\label{fig:characteristics}
\end{figure}

We chose a number of Alloy models that cover a range of
interesting characteristics: 1) number of types (\# T); 
2) maximum depth of applications (Application); 3) maximum number of connected joins
(join length); 4) maximum nesting of universal quantifiers (forall);
5) maximum nesting of existential quantifiers (exists);
6) maximum arity of a relation (arity); 7) number of total functions (\# F);
and 8) number of relations (\# R). 
We manually measured these characteristics for
each of our models.  Figure~\ref{fig:characteristics} uses a radar chart
to illustrate how our different models covered these 
characteristics.
We have eleven models that are tested for different scopes for 
a total of twenty-nine tests.  Most are not satisfiable because these
tend to be harder problems for solvers, but a few  tests have 
satisfying solutions.
These models originated from Kodkod benchmarks~\cite{ToJa2007tacas}, which
were also used to test Fortress, and we created some
additional models to ensure that we had models that contained
interesting instances of the above characteristics.

\subsection{Performance}

We ran our models for different scopes with Kodkod and
with all the twelve combinations of options we described for translation.
We immediately rejected a number of combinations due to poor
performance.
Table~\ref{tab:options} shows the combinations with interesting results
and those that we immediately rejected.  Some of these combinations with poor
performance can be explained by considering the steps in the process.
For example,
there is more quantifier expansion with relations than functions, thus
the combination of Fortress and relations has poor performance.

\floatsetup[table]{objectset=centering,capposition=top}
    \begin{table}
        \begin{tabular}{lc|c|c|c|c|c}\toprule
            &\multicolumn{3}{c}{\textbf{Functions}}&\multicolumn{3}{c}{\textbf{Relations}}
            \\\cmidrule(r){2-4}\cmidrule(r){5-7}
      &Unscoped & Fortress & SMT FMF & Unscoped & Fortress & SMT FMF\\\midrule
Typed & \ding{55} & \ding{51} & \ding{55} & \ding{51} & \ding{55} & \ding{55}\\
Untyped   & \ding{55} & \ding{55} & \ding{55} & \ding{51} & \ding{55} & \ding{51}
            \\\bottomrule
        \end{tabular}
        \caption{Translation options. \ding{51} = interesting combination, \ding{55} = rejected combination }
        \label{tab:options}
    \end{table} 
 
\begin{figure}
\includegraphics[width=\linewidth]{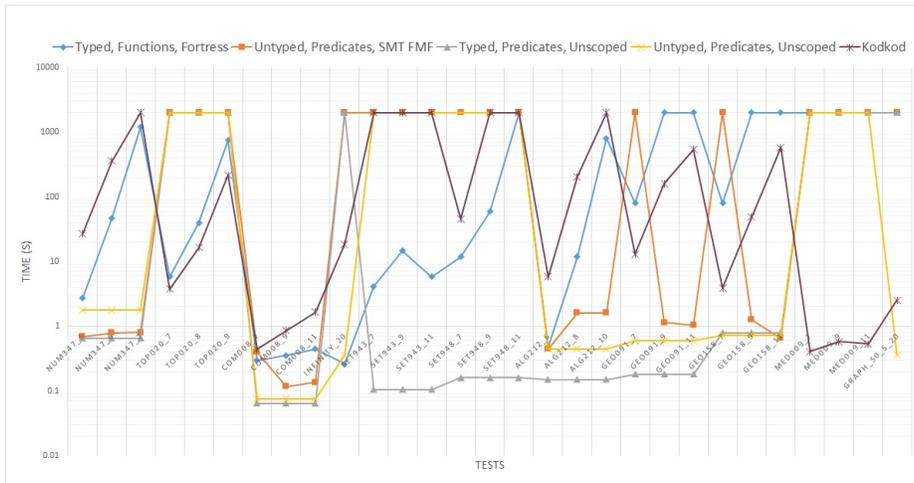}
\caption{Performance results for tests.}
\label{fig:perf_graph}
\end{figure}

All of our tests were completed on a computer with 2.6-GHz Intel Core i5 CPU, with a 2500MB Memory limit and a 2000 second time limit for 
each process. Figure~\ref{fig:perf_graph} shows our results using a logarithmic scale for time on the y-axis
with the twenty-nine tests across the x-axis.  The times include the time for translation and
the time for SMT solving (or SAT solving in Kodkod).
The five option combinations with interesting results 
are shown by different lines in the graph.  The lowest line for a test on the graph
means the best performance.  Any lines that hits the uppermost point on the graph mean that we stopped
the test after it had taken 2000 seconds or had run out of memory.

The differences
in the performance of the methods is quite considerable in most tests,
ranging from less than a second for the best performing method to over 2000 seconds
for the worst performing method.

In nine of the eleven models, one of our translation combinations produces 
better results than Kodkod.
The typed, predicates, unscoped  option had the best performance in 
five models and tied for the best
performance on three other models making it the clear winner 
in performance.
Every combination of options that we included won at least one test. 
The combination of typed, functions, Fortress performed the best for the
scoped combinations (including Kodkod).

\subsection{Analysis of Performance Results}

We hypothesized that using types
would improve the performance of the SMT solver.
When working with the Fortress option, the typed option works much
better than untyped.
However, although there are slightly better results with the typed
option for other option combinations, the results are not conclusive.

Second, we hypothesized that using predicates instead of
functions, can improve performance, at least for UNSAT problems.
Most of the UNSAT problems are solved very quickly
when predicates are used, however, they
timed out, or ran out of memory, when the functions option is used.

Third, we hypothesized that the unscoped option would often be
faster than finite scopes.   Many of the tests
with unscoped option are solved within seconds.  However, it
is unclear whether this good performance is due to the unscoped option or
the predicates option.

Although we do not have a large benchmark, we tried the linear regression method 
to find a correlation between the model characteristics and the performance of a method. 
Since the number of tests is relatively small for such method, we used it only to 
rank the characteristics, and we present a model only for the ones with an 
R-squared larger than 0.6. 
This threshold eliminated the models for Kodkod and the untyped, predicates, SMT FMF method.

The first model we present is for the option combination of typed, functions, Fortress  combination.
This model's R-squared is 0.76, and it points out that the number of types, 
the depth of function applications, and the number of relations, play the most 
important roles in its performance. 

The second model is for the typed, predicates, unscoped  combination, which has an 
R-squared of 0.65. This model points out that the number of types and the number of 
functions and relations play the biggest role in performance time of the tool. 

Lastly, the untyped, predicates, unscoped combination has the best suiting model 
with an R-squared of 0.86. This model, as expected, does not value the number of 
types as much. This model mostly emphasizes the arity of the 
functions and relations, the depth of function applications, and 
the number of functions and relations.

These results give us insight 
about which options may be more suitable for problems of certain characteristics. 
For example, we hypothesize that tests with many types may be better solved 
by the untyped option, while tests with functions of large arities may be better solved 
with other options. These insights can be used as guidelines in the Alloy modelling process 
to create models that can be solved more easily by a specific option.

Additionally, these results raise some questions regarding the internal process of 
SMT solvers.
We had some unexpected results, such as models ``Geo091'' and ``Geo158'', which could be solved 
with the untyped, predicates, SMT FMF combination easily for scopes of 9 and 11, 
but could not be solved within the 2000 seconds time limit for a scope of 7. 
Also, in model ``Set943'', the typed, functions, Fortress  combination
solved the larger scope of 11 
faster than the smaller scope of 9. 
This test was repeated multiple times, and the same result was observed each time.

\section{Related Work}

In this section, we discuss related efforts to translate Alloy to SMT-LIB.  Compared to 
previous work, we investigate and evaluate multiple options for the translation
and try to correlate them with model characteristics.  While other work has evaluated
the solving performance of their own translation, none of these works compare
solving time for unbounded with bounded scopes in SMT solvers.  
Also, we start from the Kodkod
interface for ease of future integration with the Alloy Analyzer.  Translating from
Kodkod rather than the Alloy language was easier with respect to having a more
basic language to work with, but harder because we had to reverse engineer from Kodkod
the types and functions of the Alloy model. 
We do not yet support the transitive closure operators, set cardinality or built-in types
and some of these related efforts do support these operations/types.

El Ghazi \myetal~\cite{elghazi-corr-15} describe a translation directly from Alloy to the
SMT solver Yices~\cite{yices}.
Since Yices Version 1 supported subtypes, Alloy's subtyping could be directly translated into its
Yices equivalent.
Partial functions in Alloy are translated to total functions in Yices 
by including an empty range value.  The focus of their work is on using Yices' theories
for Alloy's built-in types in order to leave these types unbounded.  They evaluate their
translation using one case study.

In El Ghazi \myetal~\cite{elghazi-fm11}, a translation directly from Alloy to Z3 is described.
It corresponds to our typed, unscoped option.  They use relations at first and then 
do some simplifications for Alloy functions. Their work supports Alloy's built-in types 
and the closure operations for relations. 
Their results shows Z3 performed
well, solving a number of problems in Alloy.  Rather than using helper functions to
translate the set expressions, they take a top-down approach to translation, passing
arguments down to the leaf relations.

Ulbrich \myetal~\cite{ulbrich-assistant} describe a translation from Alloy to the KeY
theorem prover~\cite{key} for first-order logic to check Alloy models over unbounded scopes.
Their translation matches our untyped, relations, and unscoped option.
They introduce helper relations to translate the set expressions using axioms
similar to our constraints on the helper relations.
KeY integrates automic and interactive proof and includes support for some of Alloy's
built-in types.  Their translation handles transitive closure and set cardinality, but
these may require interactive proof methods. Their results show that a number of 
Alloy assertions could be proven automatically in the KeY prover.

Reynolds et al.~\cite{ReTi13cade} propose a theory called Finite Cardinality Constraints (FCC) 
for doing finite model finding within
an SMT solver. The theory is based on the EUF subset of FOL.  Vakili and Day~\cite{VaDa16fm}
report that this method did not have good performance compared to Fortress. 

Bansal \myetal\cite{bansal-corr17} introduces a new theory for solving relational FOL (including
set cardinality) of unbounded scope problems
in SMT solvers. This is a calculus for relational logic in SMT which can be combined with their finite 
model finding feature. It has been implemented in CVC4~\cite{barret-cav11} and evaluated
on some problems but not yet linked with Alloy for evaluation.

Alloy2B\cite{alloy2b} is a tool that translates Alloy models to the B language~\cite{abrial1996}, making
a variety for B tools available for use on Alloy models including model checkers and
interactive proof tools for examining a model of unbounded scope.

We have not yet translated the transitive closure operators.  For a finite scope, it
is possible to expand the meaning of transitive closure as is done by Kodkod.
El Ghazi \myetal\cite{elghazi-nfm-15} addresses this problem for unbounded scope 
by axiomatizing  transitive closure in FOL, in an iterative manner.

\section{Conclusion}

We have presented an evaluation of various options for 
translating relational FOL as
it is represented in Kodkod to typed FOL in SMT-LIB.
We considered many options for the translation including:
typed vs untyped, predicates vs functions, and unbounded vs
bounded scopes where the formulas are either expanded pre-solving
or during SMT solving.  Our results show that with the Z3 SMT
solver, the typed, predicates, unscoped combination is the
best combination in general for unbounded scopes; and 
the typed, functions, Fortress translation combination is the 
best for bounded scopes.
There
are many interesting
directions from our work to understand how model characteristics
relate to solver performance, which could provide the basis
for a portfolio of solvers for Alloy 
(perhaps based on Why3~\cite{healy-16}). 

We have several directions for future work.  We plan to 
investigate translations for the transitive closure operator,
set cardinality, and the mapping of Alloy's built-in types
to SMT theories.  
We hypothesize that the use of SMT solvers for these built-in types
may provide better performance with unbounded scopes on these types.
Also, for a satisfiable instance, we do not yet return the instance
from the SMT solver to Alloy.  This step becomes relevant when
we integrate with the Alloy Analyzer, which is our next step.
And we would like to broaden our analysis to include more
SMT solvers or finite model finding techniques.

\section{Acknowledgments}

We thank Eunsuk Kang for help with the Alloy Analyzer and Joseph Poremba for
his work on Fortress.

\bibliographystyle{splncs04}

\end{document}